\documentclass[12pt]{iopart}
\usepackage{graphicx}
\usepackage{subfigure}
\usepackage{amssymb}

\newcommand {\snn}      {\sqrt{s_{_{\rm NN}}}}

\newcommand {\pT}		{p_{\rm T}}

\newcommand {\dphi}     {\Delta\phi}

\begin{document}
\title[Heavy-flavor particle correlations in STAR]{Heavy-flavor particle correlations in STAR
via electron azimuthal correlations with $D^{0}$ mesons}
\author{Andr\'e Mischke\footnote{This work was supported by a Veni grant from the Netherlands Organization for Scientific Research.}
for the STAR Collaboration\footnote{For the full author list and acknowledgement see Appendix "Collaboration" in this volume.}}
\address{Institute for Subatomic Physics, Utrecht University, Princetonplein 5,\\ 3584 CC Utrecht, the Netherlands.}
\ead{a.mischke@uu.nl}

\begin{abstract}
We present first STAR measurement on two heavy-flavor particle correlations in $\sqrt{s}=200$ GeV $p+p$ collisions at RHIC. 
Heavy-flavor (charm and bottom) events are identified and separated on a statistical basis through their characteristic decay topology using azimuthal correlation of non-photonic electrons and reconstructed open charmed mesons.
The results are compared to simulations from PYTHIA and MC@NLO event generators.
The gluon splitting contribution is found to be small. 
The relative bottom contribution to the non-photonic electrons is $\sim50\%$ at $\pT > 5.5$ GeV/c.
\end{abstract}


\noindent {\em 1. Introduction} \\
The study of heavy-flavor production in heavy-ion collisions provides key tests of parton energy-loss models and, thus, yields profound insight into the properties of the produced highly-dense QCD matter~\cite{Yifei}.
Theoretical models based on perturbative QCD predicted that heavy quarks should experience smaller energy loss in the medium than light quarks when propagating through the extremely dense medium due to the mass-dependent suppression (called "dead-cone effect")~\cite{El:deadcone}. 
Surprisingly, STAR measurements in central Au+Au collisions have shown~\cite{Star:npe} that the high $\pT$ yield of electrons from semi-leptonic charm and bottom decays is suppressed to the same level as observed for light-quark hadrons.
Energy-loss models describe the observed suppression reasonably well only if the bottom contribution to the non-photonic electrons is very small~\cite{Star:npe}.  
Since these  measurements are sensitive to the sum of charm and bottom decays and the crossing point of where bottom starts to dominate over charm is not accurately known~\cite{Mat05}, it is of great interest to disentangle the relative contributions experimentally.

In this paper, we present a novel analysis technique to identify and separate charm and bottom quark events via leading electron azimuthal correlations with open charm mesons. 
Requiring $e-D^0$ coincidence in the same event significantly improves the signal-to-background ratio over either technique individually. 
The shape of the azimuthal correlation distribution allows a more differential comparison between the charm and bottom contributions owing to their different decay kinematics.
%
\begin{figure}[t]
\begin{center}
\subfigure{\includegraphics[width=0.435\textwidth]{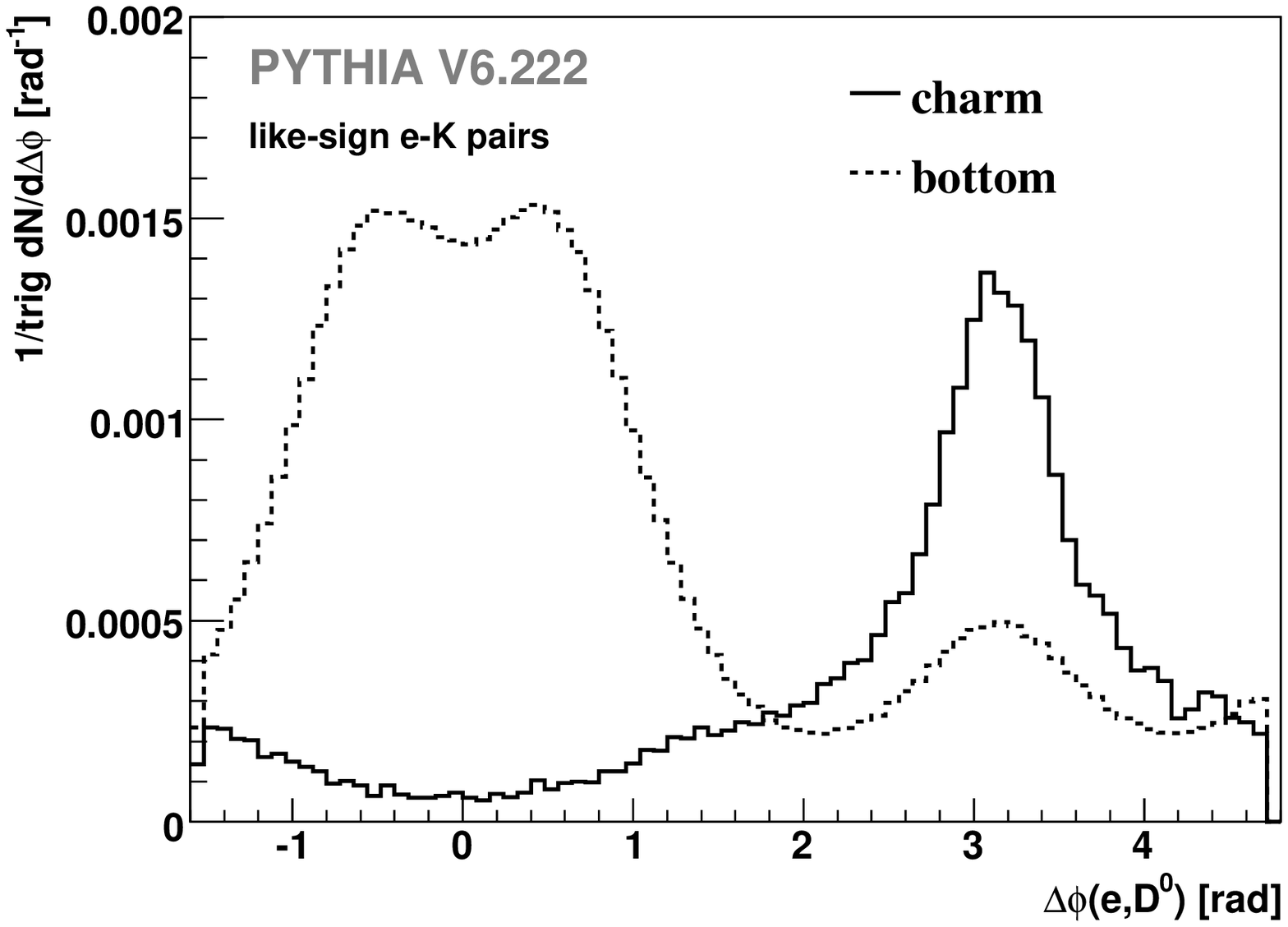}}
\hspace*{0.5cm}
\subfigure{\includegraphics[width=0.435\textwidth]{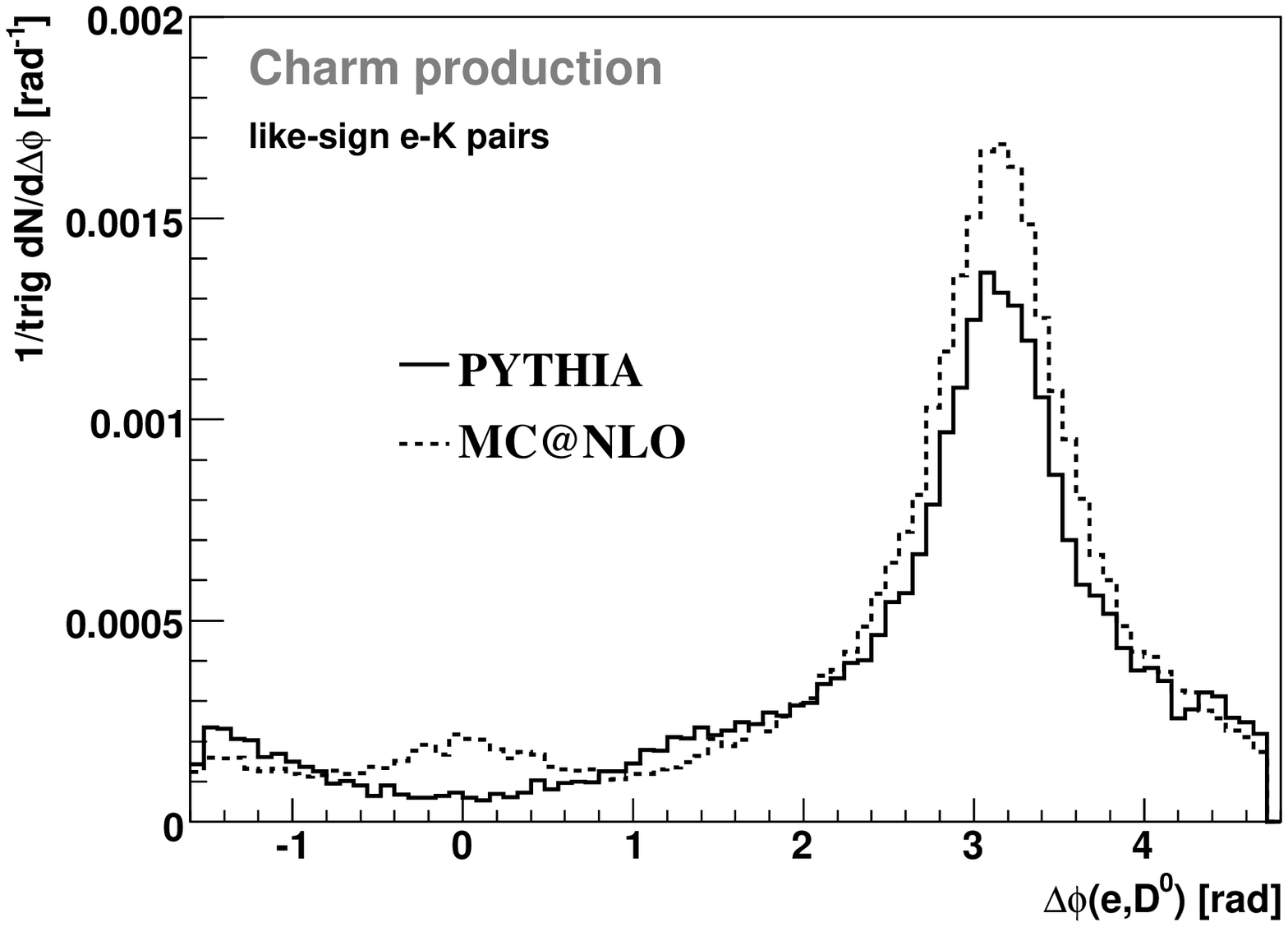}}
\vspace*{-0.5cm}
\caption[]{\label{fig:1}
Azimuthal correlation distribution of non-photonic electrons and $D^0$ mesons from charm and bottom production obtained from PYTHIA (left panel) and for charm production only from PYTHIA and MC@NLO simulations (right panel).}
\end{center}
\end{figure}

\noindent {\em 2. Correlation technique} \\
Due to flavor conservation, heavy quarks from initial hard scatterings are always produced in quark anti-quark pairs. A more detailed understanding of the underlying production process can be obtained from events in which both heavy-quark particles are detected. Momentum conservation implies that these heavy-quark pairs are correlated in relative azimuth ($\dphi$) in the plane perpendicular to the colliding beams.
This correlation survives the fragmentation process to a large extent in $p+p$ collisions. 
In this analysis, charm and bottom production events are identified using the characteristic decay topology of their jets. 
Charm quarks predominantly hadronize directly (BR = $56.5\pm3.2\%$) and bottom quarks via $B$ decays into $D^0$ mesons. 
The branching ratio for charm and bottom quark decays into electrons is 9.6$\%$ and 10.86$\%$, respectively~\cite{Eid04}.
While triggering on the leading electron (trigger side), the balancing heavy quark identified by the $D^0$ meson can be used to determine the underlying production mechanism (probe side). 
A charge-sign condition on the trigger electron and decay Kaon is applied to separate charm and bottom quark events. 
The azimuthal correlation distribution was studied using PYTHIA simulations~\cite{mySQM07}. 
It has been shown that $e-K$ pairs with the same charge sign (called like-sign $e-K$ pairs) exhibit a near-side peak from $B$ decays whereas the away-side peak is dominated by charm pair production (cf. Fig.~\ref{fig:1}, left panel). 
The complete NLO contributions (including gluon-splitting diagrams) were estimated using MC@NLO simulations, a QCD computation with a realistic parton shower model~\cite{Mod:frix}. 
The comparison of the near-side correlation yield beween PYTHIA and MC@NLO (cf. Fig.~\ref{fig:1}, right panel) indicates that gluon splitting contributes $\sim5\%$ of the open charm production observed at RHIC, consistent with the STAR measurment of the $D^*$ content in jets~\cite{Tim}. \\
\begin{figure}[t]
\begin{center}
\subfigure{\includegraphics[width=0.41\textwidth]{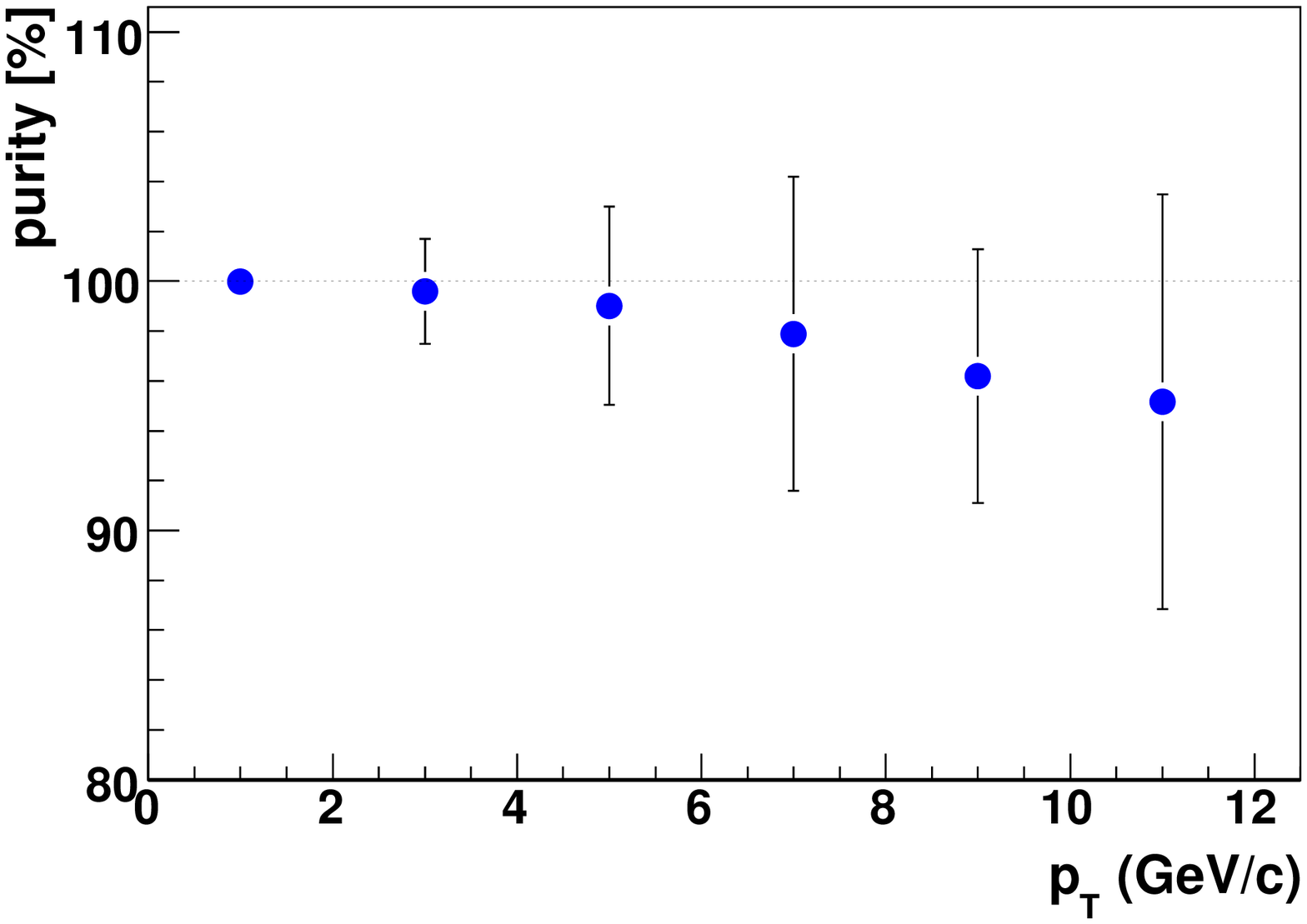}}
\hspace*{0.5cm}
\subfigure{\includegraphics[width=0.45\textwidth]{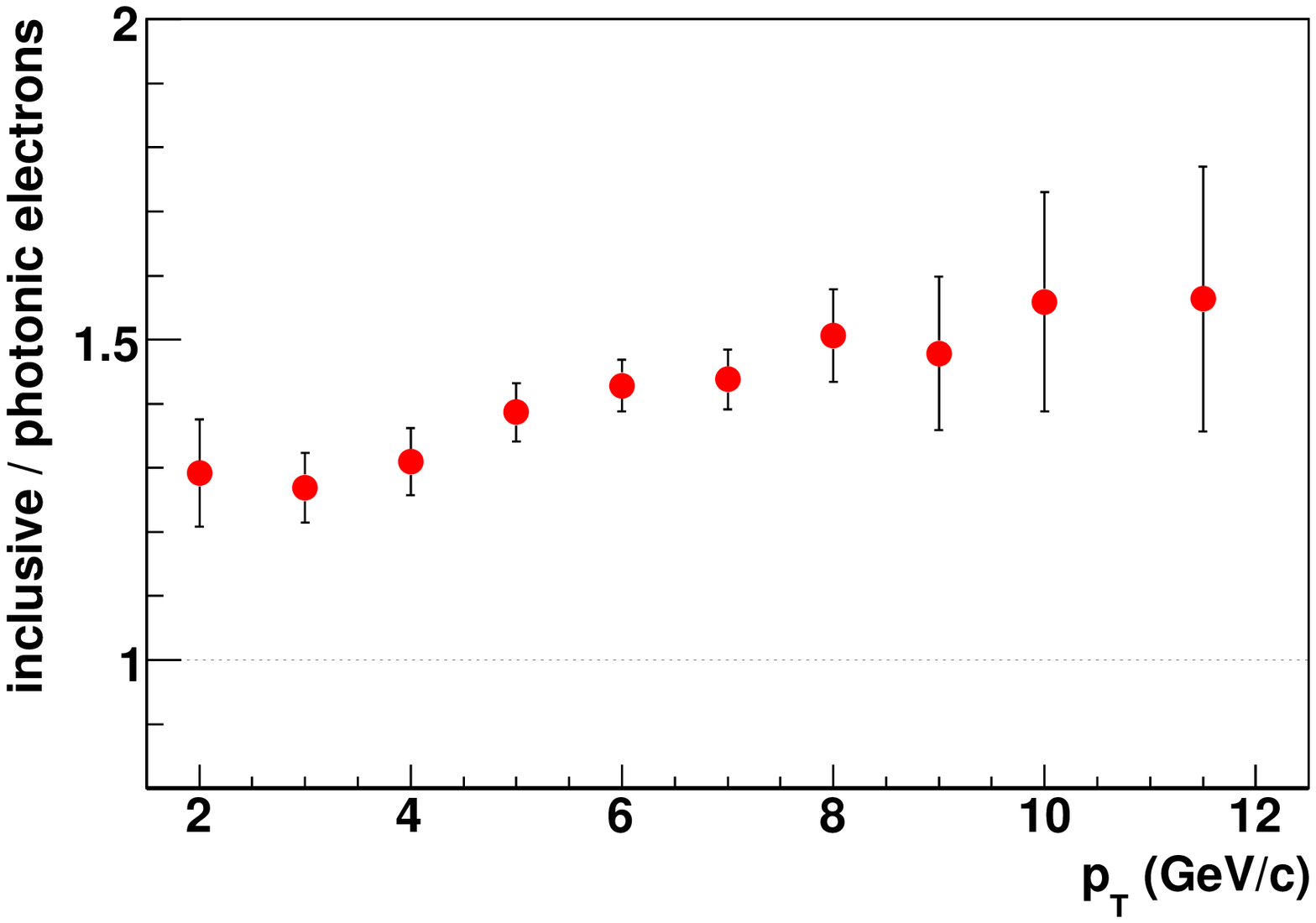}}
\vspace*{-0.5cm}
\caption[]{\label{fig:2}
Electron purity (left panel) and the ratio of inclusive electrons to photonic background as a function of trigger-electron $\pT$ (right panel).}
\end{center}
\end{figure}

\noindent {\em 3. Data analysis} \\
The analysis is performed using Run VI $p+p$ data taken at $\sqrt{s}=200$ GeV with an integrated luminosity of $\sim 9$ pb$^{-1}$.
The amount of material within the detector volume causing photon conversions is minimized by a tight collision vertex cut, $|z|~<~30$~cm.
Charged particles are identified by their ionization energy loss in the Time Projection Chamber (TPC), which also provides tracking over a large kinematical range ($|\eta|<1.4$ and full azimuth) with excellent momentum resolution.
The STAR detector~\cite{Star:det} utilizes an electromagnetic calorimeter (BEMC) as a leading electron or photon trigger to study high $\pT$ particle production. 
A high-tower trigger with an energy threshold of 5.4~GeV for the highest energy ($E_T$) in a BEMC cell was used to enhance the high $\pT$ range. 
The shower maximum detector
measures the profile of an electromagnetic shower and the position of the shower maximum with high resolution. 
Electron identification is performed by combining the information from the TPC and the BEMC (cell energy). 
A cut on the shower profile size combined with a requirement on the ratio momentum-to-cell energy, $0<p/E<2$, and a momentum dependent cut on the ionization energy loss, $3.5<dE/dx<5.0$~keV/cm, reject a large amount of hadrons (hadron suppression factor $10^{3}-10^{5}$). 
Background electrons from photon conversions and Dalitz decays are identified and rejected based on invariant mass~\cite{Star:npe}. 
The resulting electron purity and ratio of inclusive electrons to photonic background are illustrated in Fig.~\ref{fig:2}, left and right panel, respectively.
$D^0$ mesons are typically reconstructed via their hadronic decay $D^0 \rightarrow K^- \pi^+$ (BR = 3.84$\%$) by calculating the invariant mass of all oppositely charged TPC tracks in the same event, where negative tracks have to fulfill a $dE/dx$ cut of $\pm 3\sigma$ around the Kaon band to enhance the Kaon candidate probability. 
Only events with a non-photonic electron trigger are used for the $D^0$ reconstruction which suppresses the combinatorial background significantly. 
The resulting invariant mass distribution of $K-\pi$ pairs shows a pronounced $D^0$ peak around the expected value~\cite{Eid04} with a signal-to-background ratio of 14$\%$ and a signal significance of 3.7~\cite{mySQM07}. \\ 
\begin{figure}[t]
\begin{center}
\subfigure{\includegraphics[width=0.45\textwidth]{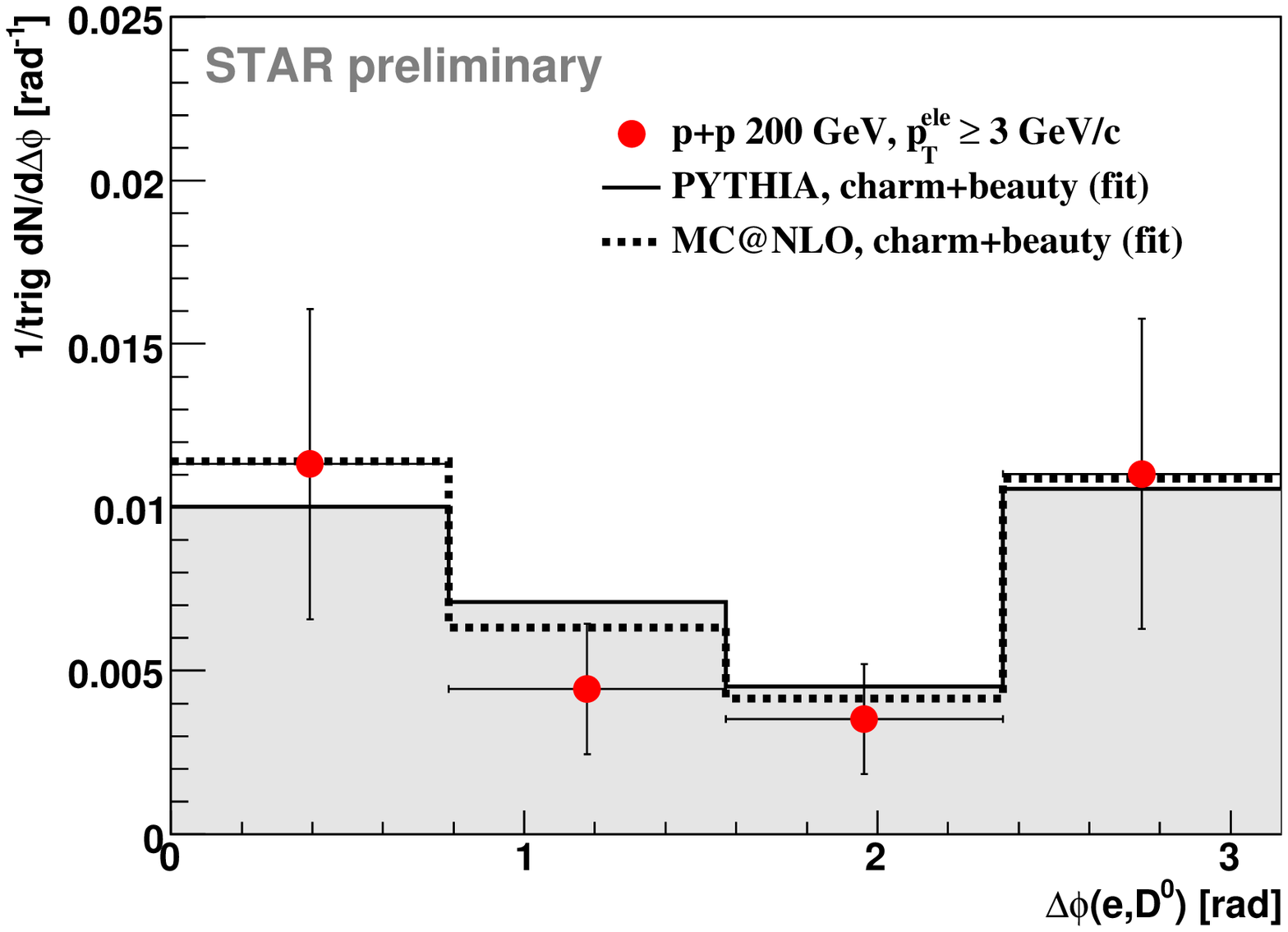}}
\hspace*{0.3cm}
\subfigure{\includegraphics[width=0.46\textwidth]{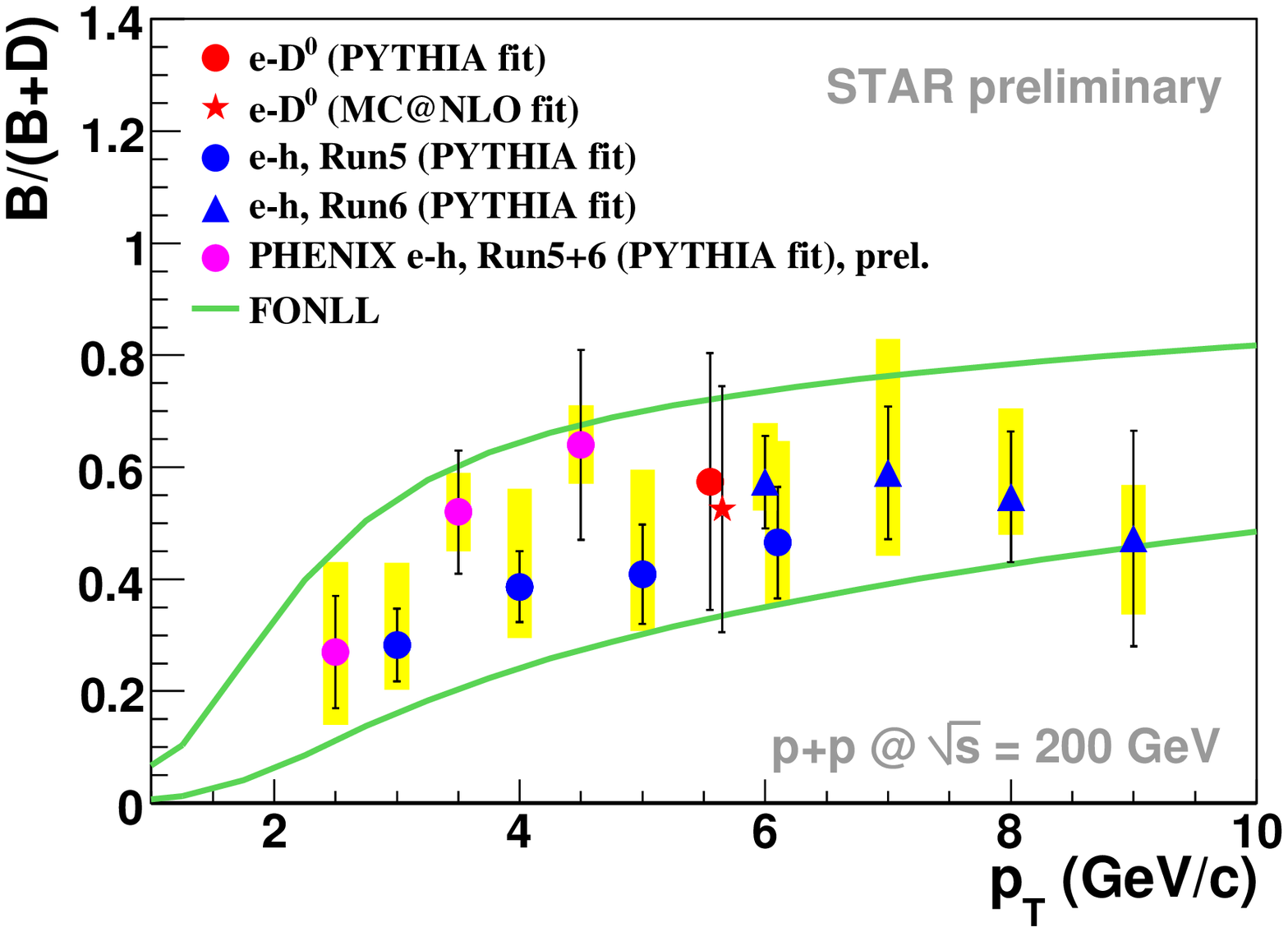}}
\vspace*{-0.5cm}
\caption[]{\label{fig:3}
Left panel: Azimuthal correlation distribution of non-photonic electrons and $D^0$ mesons (for like-sign $e-K$ pairs) in $p+p$ reactions at $\snn$~=~200 GeV. Statistical errors are shown only. The grey histogram (dashed line) illustrates results from PYTHIA (MC@NLO) simulations.
Right panel: Relative bottom contribution to the total non-photonic electron yield derived from e$-D^{0}$ and e$-$h correlations~\cite{Gang, Morino} and compared to the uncertainty band from a FONLL calculation.}
\end{center}
\end{figure}

\noindent {\em 4. Results and conclusions} \\
The azimuthal correlation distribution of non-photonic electrons and $D^0$ mesons is shown in Fig.~\ref{fig:3}, left panel, which exhibits a clear near- and away-side correlation peak with similar yields.
The results are compared to PYTHIA and MC@NLO simulations.
The observed away-side correlation peak can be attributed to prompt charm pair production ($\sim75\%$) and $B$ decays ($\sim25\%$).
By contrast, the near-side peak represents essentially contributions from $B$ decays. 
The relative bottom contribution to the non-photonic electrons has been extracted. 
The results are shown in Fig.~\ref{fig:3}, right panel, together with the results from electron$-$hadron azimuthal correlations~\cite{Gang, Morino}. 
These data provide convincing evidence that bottom contributes significantly to the non-photonic electron yields at high $\pT$. 
For example, the comparison between the data and the MC@NLO simulation suggests that $B/(B+D) = 52\pm21\%$ for $\pT > 5.5$ GeV/c.
Combining these results with the nuclear modification factor $R_{AA}$ observed for non-photonic electrons in central Au+Au collisions~\cite{Star:npe}, this suggests substantial suppression of bottom production at high $\pT$ in the produced medium which contradicts expectations based on pQCD inspired models of parton energy loss.
This correlation method in combination with the STAR detector upgrades will allow comprehensive energy-loss measurements of heavy quarks in heavy-ion collisions in the near future. \\


\end{document}